\documentclass[]{interact}
\usepackage{epstopdf}
\usepackage[caption=false]{subfig}

\usepackage{comment}
\usepackage[numbers,sort&compress]{natbib}

\theoremstyle{plain}

\theoremstyle{definition}

\theoremstyle{remark}

\usepackage{url}
\usepackage{soul}
\usepackage{xcolor} 

\sethlcolor{yellow}

\begin{document}

\title{Bioscience students in physics courses with higher test anxiety have lower grades on high-stakes assessments and women report more test anxiety than men}

\author{
  Alysa Malespina,
  Fargol Seifollahi$^{*}$\thanks{$^{*}$Corresponding author. Email: fas102@pitt.edu},
  Chandralekha Singh\\
  \textit{Department of Physics \& Astronomy, University of Pittsburgh, Pittsburgh, PA, USA 15260}
}

\maketitle

\begin{abstract}

Test anxiety is beginning to be recognized as a significant factor affecting student performance in science, technology, engineering, and mathematics (STEM) courses, potentially contributing to gender inequity within these fields. Additionally, the management of test anxiety can improve self-efficacy, which is a construct that has been well studied in the physics context.  In this study, we investigated the relationship between self-efficacy, test anxiety, and gender differences in performance in a two-semester-long introductory physics course sequence for bioscience students in which women outnumber men. 
Using validated survey data and grade information from students in a two-semester introductory physics course sequence, we compared the predictive power of self-efficacy and test anxiety on female and male students' performance on both low- and high-stakes assessments. We found that there were gender differences disadvantaging women in self-efficacy and test anxiety in both Physics 1 and Physics 2, as well as gender differences in high-stakes outcomes in Physics 1. 
There were no gender differences in low-stakes assessment scores. We also found that self-efficacy and test anxiety predicted high-stakes (but not low-stakes) assessment outcomes in both Physics 1 and Physics 2.  
Comparison of these findings with prior studies involving physical science and engineering students shows that although women outnumber men in physics courses for bioscience students and the career goals of bioscience students are very different from the earlier researched group, most of the negative trends hold even for this new population. 
Thus, these findings, in a new context involving bioscience students in physics courses, are very important because they reinforce the systemic nature of women being affected more adversely by anxiety in high-stakes assessments, which is a threat to creating equitable and inclusive learning environments. 
An important implication is that course instructors should carefully consider how high-stakes and low-stakes assessments are used to determine grades and how to create an overall equitable, inclusive, and low-anxiety learning environment.

\end{abstract}


\section{Introduction and Theoretical Framework}

Students' science, technology, engineering, and~mathematics (STEM)-related motivational beliefs have implications for their performance in individual courses as well as their long-term outcomes~\cite{bouffardbouchard1991, pintrich1990, zimmerman2000, nissen2016, sawtelle2012}. In~particular, students' motivational beliefs are correlated with their goals and tend to predict student learning outcomes~\cite{bouffardbouchard1991, pintrich1990, steele1997threat, zimmerman2000, maries2024towards, nissen2016, sawtelle2012}. There also tend to be differences between men's and women's motivational beliefs regarding physics, which have been linked to performance differences in physics courses~\cite{salehifrontiers, marshman2018femalewithA, cwik2022gender, nissen2016, cavallo2004}. Here, we focus on two motivational factors: test anxiety and self-efficacy.  Test anxiety can affect students’ test performance and is more likely to affect women~\cite{zeidner1998}. Self-efficacy in a given domain is a student’s belief in their ability to succeed at an activity or subject or complete a task~\cite{bandura1999}. 

Test anxiety can impact students' cognition, physical body, and~behavior~\cite{zeidner1998, maloney2014anxiety}.  When they experience test anxiety, students' cognitive resources are not entirely  devoted to the assessment, but~can be taken up by worry and intrusive thoughts of failure~\cite{zeidner1998}. Additionally, test anxiety can affect how students feel during an assessment. For~example, they may experience a fast heartbeat or ``butterflies in their stomach''.  The~behavioral aspect of test anxiety manifests in avoidance techniques, such as procrastination or interacting only with surface-level feedback after the exam (e.g., not examining mistakes closely to make a plan for future improvement)~\cite{zeidner1998, solomon1996}. Other studies have found that test anxiety negatively affects student performance, especially on high-stakes assessments such as exams that are heavily weighted~\cite{ballen2017a, stang2020, malespina2022calc_TA}. In~addition,  women are more likely to report test anxiety than men~\cite{zeidner1998, bandura1999}, so understanding test anxiety and how to minimize its effect on student success is vital in creating positive and equitable learning~environments.

Self-efficacy~\cite{bandura1999, bandura_2012} has been linked to positive learning outcomes for physics students~\cite{cavallo2004, nissen2016, maries2024towards, sawtelle2012}. Self-efficacy of students in a particular domain can be enhanced through several mechanisms. One way is by overcoming difficulties, such as succeeding in a challenging homework assignment~\cite{bandura_2012}. Self-efficacy can also be formed through social means, for~example, through observation of role models succeeding in the domain of interest or~through receiving encouragement such that it allows for students to measure their success through personal improvement~\cite{bandura_2012}. The~final mechanism is regulation of emotional states, such as management of anxiety~\cite{bandura_2012}.
Women may have fewer role models due to under-representation of women in physics~\cite{porter2019, hazari2017importance}, and~they are less likely to receive encouragement that they can succeed in physics from their instructors and peers~\cite{li2023impact, sundstrom2022introductory}.  Because~high self-efficacy allows students to develop coping mechanisms that could reduce test anxiety, we hypothesize that students with high self-efficacy are also likely to have low test anxiety~\cite{bandura1999}. 

In this research, we aim to investigate whether test anxiety and/or self-efficacy  predict low- and high-stakes assessment outcomes for women and men in a novel context: a two-semester physics course sequence for bioscience students in which women outnumber men. Here, low-stakes assessments are those that individually make up a small portion of a student’s grade, such as homework. High-stakes assessments are individual assessments that make up a large portion of a student’s grade such as traditional exams~\cite{ballen2017a, cotner2020, pelch2018}. For~example, while weekly homework assignments that add up to 10\% of the total grade (with each of them counting toward less than one percent of the grade) are considered as low-stakes assessments, a~heavily weighted final exam that individually accounts for 25\% of the total grade is considered as a high-stakes assessment. Prior research in a physics course for physical science and engineering students and students in other STEM courses  has shown that gender gaps are more prevalent in high-stakes than low-stakes assessments~\cite{ballen2017a, cotner2020, pelch2018}. This, combined with gender differences in physics self-efficacy and test anxiety among physical science and engineering students~\cite{malespina2022calc_TA}, led us to hypothesize that physics test anxiety and self-efficacy may predict high-stakes but not low-stakes physics assessment performance and may more adversely affect women even in physics courses in which they outnumber men. Prior research has found that there is a relationship between test anxiety and high-stakes assessment grades in biology classrooms~\cite{ballen2017a}. Additionally, past research shows a relationship between self-efficacy, test anxiety, and~high-stakes physics grades as well as gender differences disadvantaging women for students enrolled in introductory physics courses for engineering and physical science majors in which women are under-represented~\cite{malespina2022calc_TA}. Past research shows that in general, bioscience students differ from physical science and engineering majors in both interests and motivations~\cite{couch2023_incollection}. 
Therefore, it is very important to investigate how systemic these inequitable trends are across different contexts, e.g.,~for an entirely different student population within a given STEM discipline such as~physics. 

It is important to investigate whether the numerical under-representation of women in some physics courses is the main factor contributing to gender differences. For~example, in~physics, there are certain stereotypes about who can excel in the discipline and that one needs to be a genius to do well, which may disproportionately affect women and lead to greater anxiety than men even in physics courses in which they outnumber men. Thus, here we focus on these issues related to physics test anxiety and self-efficacy in high-stakes and low-stakes assessment for students who have never been the focus of this type of investigation in the past: female and male students in introductory physics for bioscience and health science-related majors where women are in the majority. Comparison of our findings, discussed in detail in the Results Section, with prior studies involving physical science and engineering students shows that although women outnumber men in physics courses for bioscience students and the career goals of bioscience students are very different from the earlier researched group, the~trends hold even for this new population~\cite{malespina2022calc_TA, Malespinadissertation, Malespina2024calc_TA}.
Thus, the~findings discussed here in the context of physics courses, involving bioscience students and other students with interest in health-related majors, are very important because they emphasize the 
deep-rooted nature of 
women being affected more adversely by anxiety in high-stakes assessments, which is a major impediment to creating equitable and inclusive learning~environments.

Students pursuing bioscience and health science-related majors are generally required to take at least one physics course for their major (and many of them are required to take two physics courses). Women are not under-represented in these physics courses for bioscience and health science-related majors, but~there may still be a gender gap in the motivational belief scores of students in the course.  In~particular, prior research has found that even in physics courses in which women are not under-represented, men tend to have higher grades and physics-specific motivational beliefs than women~\cite{little2019, espinosa2019, wang2018, hazari2018,  brewe2010, traxler2015, vandusen2020_equity,cwik2022gender}. For~example, women tend to have lower physics self-efficacy than men with the same grades in courses for engineering and physical science students as well as courses for students with interest in bioscience and health science-related professions~\cite{marshman2018femalewithA,cwik2021damage}.

Our goals for this research were to investigate the relationship between test anxiety and assessment outcomes in introductory physics courses, with~a focus on gender differences in each construct, in~this novel context of students in bioscience and health science-related majors.  We hypothesized that test anxiety would predict high-stakes but not low-stakes assessment outcomes among students. We also hypothesized that despite women outnumbering men in these physics courses, women would still be more likely to experience higher levels of test anxiety partly due to the stereotype that physics requires innate genius and~because women tend to stay away from disciplines linked to exceptional innate ability~\cite{leslie2015}. 
We included self-efficacy in our investigation because the relationship between performance and self-efficacy is well documented in physics~\cite{whitcomb2020comparison, cavallo2004} and~because management of anxiety is explicitly mentioned as a mechanism to enhance self-efficacy~\cite{bandura_2012}.

\subsection{Research Questions}

 With this background and goals in mind, we aim to answer the following research questions in the novel context of a two-semester physics course sequence for bioscience students and students interested in health professions: 

\begin{enumerate}
    \item[RQ1.] Are there gender differences in students’ prior preparation, self-efficacy, or~test anxiety?
    \item[RQ2.] Are there gender differences in students’ low- or high-stakes assessment scores?
    \item[RQ3.] Are there gender differences in terms of low- or high-stakes assessment scores for students with the same self-efficacy and/or test anxiety?
\end{enumerate}

\section{Methodology}

\subsection{Participants and Procedures}

This study took place at a large research university in the United States. Participants were students enrolled in a Physics 1 or 2 course for bioscience and other health-related majors. Intended majors for students in our sample, aside from biological sciences, include but are not limited to microbiology, neuroscience, molecular biology, chemistry with bioscience option, etc., where the common theme across these students is their interest in health-related professions. Previous research has shown that for those aiming for health-related careers and enrolled in science courses, there are various pathways that can lead to or away from their initially intended majors~\cite{witherspoon2020locating}. Therefore, the~many minor distinctions within these majors make it challenging to analyze them separately with sufficient statistical power. We will use the term ``bioscience majors" throughout for all students since they comprise the majority of students in both Physics 1 and Physics 2 and both these courses are mandatory for them. Students can only advance to Physics 2 if they have at least a C grade in Physics 1, equivalent to a 2.0 out of a 4.0 point grade scale at this institution, with~3.0 corresponding to a B grade and 4.0 corresponding to an A grade. Some of the low-enrollment health-related majors only require Physics 1, so students may not take Physics 2 unless they want to take the MCAT exam for medical school~admission. 

The Physics 1 course primarily covered mechanics, though~both thermodynamics and waves were also included. The~Physics 2 course covered electricity and magnetism, geometrical optics, and~physical optics. The~courses included 2 weekly sessions of 75~min traditional lecture-based instruction led by the course instructors, along with smaller-sized 50 min recitation sessions taught by teaching assistants in which students worked collaboratively on solving physics problems. These courses did not include a laboratory component, although~many of the students taking them end up taking laboratory courses as an elective later. The~Physics 1 and Physics 2 student samples included sections taught by the same instructor, although~the students were not necessarily the same across the two courses. This helped us ensure that no instructor-level effects went unaccounted for in our models. For~both courses, midterm and final exams comprised 40\% and 25\% of the final course grade, respectively. We considered the midterm and final exams as high-stakes assessment as they were each highly weighted and altogether made up approximately two-thirds of the final course grade. Another 10\% of the course grade was determined by homework problems which were assigned weekly for students to complete at their own time, and~the remaining 25\% were participation grades based on completeness for students responding to clicker questions in the lectures or their work in groups on problems in the recitation sessions. Generally, we consider homework and participation as low-stakes assessments as each of them individually take up a small portion of the total grade. For~the analysis in this paper, however, we only used students' homework grades as the low-stakes outcome, as those were the only low-stakes assessments that were graded for correctness and not completeness. Throughout this paper, we use the terms ``test'' and ``exam'' interchangeably.
All the midterm exams and final exams for the Physics 1 and Physics 2 courses were supervised, timed, and in-person, and~while we mainly associate the exams being high-stakes with the grading weight, we recognize that these elements can also contribute to the anxiety levels in students. Midterm and final exams were interspersed uniformly throughout the semester, with~midterm exams held around the one-, two-, and~three-month marks after the start of classes. All exams were mainly multiple choice, with~1--3 open-ended~questions.

Students were given extra credit as an incentive for taking the survey. The~surveys were given during the first and last week of classes in the mandatory teaching assistant-led recitations. We call the first and final data sets ``pre" and ``post", respectively. In~Physics 1, 204 students took the pre-test and 210 took the post-test. In~Physics 2, 185 students took the pre-test and 89 took the post-test. The~post-tests were taken in the final week of the classes, before~students took their final~exam.

For analysis, we included only students who successfully passed an attention check on the survey (a question that requested the students to select option ``C"). Additionally, we included as many students as possible in each part of the analysis. For~example, in~a model that uses the average of one construct as well as students' standardized test scores, we would exclude students who were missing SAT and ACT scores or~were missing either pre- or post-survey results. 
One Physics 2 class section was not able to complete the post-survey and were missing post-test anxiety and self-efficacy data. This resulted in a smaller sample size for post-test anxiety and self-efficacy, but~students in this section had statistically indistinguishable prior preparation, pre-motivational factors, and~assessment outcomes from other students in the sample, so they were included in analysis where~possible.  

This research was carried out in accordance with the principles outlined in this institution’s Institutional Review Board ethical policy, and~de-identified demographic data were provided through university records. For~some variables, such as high school GPA, this approach allows us to rely on records that may be more accurate than students’ own recollection. However, it limits other measures such as student gender, for~which students could only report either “male” or ``female''. We acknowledge the potential limitations and harms that this method of data collection may cause~\cite{traxler2016}. This institution recently began to implement more inclusive gender reporting methods for students, which are planned to be used once student samples are large enough to be meaningful for quantitative analysis.  Demographic data indicated our Physics 1 sample was 67\% women and our Physics 2 sample was 56\% women. Students in Physics 1 identified with the following races/ethnicities: 66\% White, 17\% Asian, 7\% African American/Black, 6\% multiracial, 3\% Hispanic/Latinx, and~1\% unspecified.  Students in Physics 2 identified with the following races/ethnicities: 63\% White, 21\% Asian, 6\% African American/Black, 6\% multiracial, 3\% Hispanic/Latinx, and~1\% unspecified.

\subsection{Measures}\subsubsection{Self-Efficacy and Test Anxiety}

\begin{table}[tb]
\tbl{Items included in student survey and factor analysis. \label{survey}}
{\begin{tabular}{llcccc}
\toprule%
& & \multicolumn{4}{@{}c@{}}{Factor Loading}  \\
&  & \multicolumn{2}{@{}c@{}}{Physics 1} & \multicolumn{2}{@{}c@{}}{Physics 2}\\
 & \multicolumn{1}{@{}c@{}}{Construct Name/Item Text} & Pre & Post & Pre & Post \\
\midrule
& \multicolumn{1}{@{}c@{}}{Self-Efficacy} & &  \\ \cmidrule{1-2}
1. & I am able to help my classmates with physics in the laboratory or in recitation & 0.61 & 0.56 & 0.71 & 0.74  \\
2. & I understand concepts I have studied in physics & 0.60 & 0.73 & 0.76 & 0.80 \\
3. & If I study, I will do well on a physics test & 0.64 & 0.84 & 0.80 & 0.82 \\
4. & If I encounter a setback in a physics exam, I can overcome it & 0.64 & 0.79 & 0.74 & 0.83 \\
\midrule
  & \multicolumn{1}{@{}c@{}}{Test Anxiety} & & && \\\cmidrule{1-2}
5. & I am so nervous during a physics test that I cannot remember what I have learned & 0.80 & 0.83 & 0.91 & 0.89\\
6. & I have an uneasy, upset feeling when I take a physics test & 0.90 & 0.92 & 0.91 & 0.93 \\
7. & I worry a great deal about physics tests & 0.80 & 0.89 & 0.83 & 0.82 \\
8. & When I take a physics test, I think about how poorly I am doing & 0.76 & 0.81 & 0.87 & 0.92\\
\bottomrule
\end{tabular} } \bigskip
\raggedright
\small\textit{Note.} The same items were given to students for the pre and post survey.  For Physics 1, $N=230$ and for Physics 2, $N=203$.
\end{table}

All test anxiety and self-efficacy survey items can be found in Table~\ref{survey}. The~test anxiety survey questions were adapted 
from the previously validated Motivated Strategies for Learning Questionnaire~\cite{pintrich1991, pintrich_valid}. To~ensure we were measuring domain-specific constructs, we explicitly mentioned physics in the survey items, as~seen in Table~\ref{survey}. For~example, ``I have an uneasy, upset feeling when I take an exam" became ``I have an uneasy, upset feeling when I take a physics test". Self-efficacy survey questions were constructed from other surveys and were previously validated~\cite{kalender2019gendered}. Test anxiety items were on a five-point Likert scale (1---Not at all true, 2---A little true, 3---Somewhat true, 4---Mostly true, \mbox{5---Completely} true),  and self-efficacy items were on a four-point Likert scale (1---NO!, 2---no, 3---yes, \mbox{4---YES!}). All responses were placed on a 0-1 scale to account for multiple Likert scales. Higher scores on the test anxiety items indicate higher test anxiety levels; therefore, an~ideal course outcome would be high self-efficacy and low test anxiety scores on the surveys. We note that since students take the pre-surveys in the first week of their classes, their responses are based on anticipation for the course and reflect any prior physics course experience they have had. We further validated the survey through twenty one-hour student interviews to ensure that students interpreted questions as intended. These validations were incorporated early in the design of the survey before we started using the survey at this institution and~included students all the way from introductory to upper-level physics~courses.

Additionally, we performed confirmatory factor analysis using the student sample in this study as a check for continued validity. For~both the pre- and post-surveys, the~Comparative Fit Index (CFI) and Tucker Lewis Index (TLI) were $\ge$0.90~\cite{hu1999}, and~the Root Mean Square Error of Approximation (RMSEA) and Standardized Root Mean Square Residual (SRMR) were both $\le$0.08~\cite{browne1993}, which can be seen in Table~\ref{fit_indicies} in the Appendix. Cronbach’s $\alpha$ was also consistently above 0.7, indicating good internal consistency across our measures~\cite{frey2018}. Standardized factor loadings were all above 0.5~\cite{hu1999}. The~square of the standardized factor loadings gives the percentage of variance of each observed variable that is explained by the latent variable, meaning that at least 25\% of the variance in each of the survey items is explained by the respective~construct.

\subsubsection{Prior academic preparation}

High school Grade Point Average (HS GPA) was reported using the weighted 0–5 scale, which is based on the standard 0 (failing)–4 (A) scale with adjustments for honors, Advanced Placement, and International Baccalaureate courses (these programs may offer a bonus 
 as a reward for taking advanced courses, which can allow a GPA higher than 4.0). High school GPA is taken as a measure of general academic skills and generally is a strong predictor of early undergraduate course performance~\cite{galla2019}. 

Students’ Scholastic Achievement Test Math (SAT Math) scores are on a scale of 200–800 and are used as a predictor of performance on high-stakes assessments involving mathematical problem solving (e.g., physics exams) \cite{hazari2007, vincentruz2017, galla2019}. If~a student took the American College Testing (ACT) examination, we converted ACT to SAT scores~\cite{collegeboard2018}. If~a student took a test more than once, the~school provided the highest section-level score for the SAT and the highest composite score for the ACT. If~a student took both ACT and SAT tests, we used their SAT~score.

\subsubsection{Assessment Scores} Homework and exam grades were provided by the instructor and were de-identified by an honest broker before being included in analysis. If~grades were not on a 0--100 scale, they were rescaled. For~example, if~homework was graded on a 10-point scale, all scores were multiplied by 10 for~analysis. 

\subsection{Analysis}

First, we report means and standard deviations of each variable separately for men and women.  Next, to~determine if there were sex differences in the means of self-efficacy, test anxiety, prior preparation, or~assessment scores, we performed unpaired \textit{t}-tests to measure the statistical significance of the differences~\cite{frey2018} and Cohen's \textit{d} to measure the size of the difference~\cite{cohen1988}. Cohen's \textit{d} is calculated using:

\begin{equation*}
    d = \frac{\mu_{1}-\mu_{2}}{\sqrt{(\sigma_{1}^{2} + {\sigma_{2}^{2}})/2}},
\end{equation*}

\noindent where $\mu_{1}$ and $\mu_{2}$ are the mean values of each group and $\sigma_{1}$ and $\sigma_{2}$ are the standard deviations of each group~\cite{cohen1988}. Group one was women and group two was men.  Cohen’s \textit{d} is considered small if $d\sim0.2$, medium if $d\sim0.5$, and~large if $d\sim0.8$ \cite{cohen1988}. We performed this analysis separately for Physics 1 and Physics 2~courses. 

To explore the predictive relationships between test anxiety and assessment outcomes, we used multiple regression analysis. For~each regression model, we report the standardized $\beta$ coefficients, sample size, and~adjusted R-squared. Standardized coefficients were used because they are in units of standard deviation and allow for direct comparison of effects~\cite{cohen2003}. We initially used gender, SAT Math scores, and~HS GPA as predictors for low- and high-stakes assessment scores.  Here, low-stakes assessment scores are the students’ average homework grades. High-stakes assessment scores are weighted so that 75\% of the category is midterm exam grades and 25\% is the final exam grade. This weighting was performed because the instructor gave three midterm exams and one final~exam. 

After establishing baseline models, we introduced pre- or average test anxiety and self-efficacy as predictors. 
Average test anxiety/self-efficacy is the mean of pre- and post-scores and~was used as a proxy for students’ test anxiety/self-efficacy  while they were taking the course. For~both courses, we introduced two models using pre-self-efficacy and pre-test anxiety scores and~four models using the average scores. The~first two models predicted high-stakes assessment scores using both or neither of the constructs, while the third and fourth models used either self-efficacy or test anxiety in addition to the baseline predictors. During~regression analysis, we used combined assessment categories (e.g., low- and high-stakes assessments), but~results were similar when the categories were separated. For~example, the~regression models predicting high-stakes assessment scores were similar to both the models predicting midterm exam grades and those predicting final exam~grades.

\section{Results and Discussion\label{results}}
 
\subsection{RQ1.  Are There Gender Differences in Students’ Prior Preparation, Self-Efficacy, or~Test Anxiety? \label{rq1}}

\begin{table}[tb]
\begin{center}
\begin{minipage}{\textwidth}
\tbl{\raggedright Physics 1 prior preparation, self-efficacy, test anxiety, and assessment scores \label{110_table}}
{\begin{tabular*}{\textwidth}{@{\extracolsep{\fill}}lccccccccc@{\extracolsep{\fill}}}
\toprule%
& \multicolumn{3}{@{}c@{}}{Female} & \multicolumn{3}{@{}c@{}}{Male} & \multicolumn{3}{@{}c@{}}{Comparison} \\
\cmidrule{2-4}\cmidrule{5-7}\cmidrule{8-10}%
Variable & N & Mean & SD & N & Mean & SD & \textit{t} & $p$ & $d$\\
\midrule
HS GPA & 153 & 4.14 & 0.40 & 77 & 3.96 & 0.64 & 2.54 & 0.011 & 0.36  \\
SAT Math & 152 & 666 & 64 & 75 & 677 & 70 & -1.23 & 0.220 & -0.17   \\ \midrule
Self-Efficacy Pre & 135 & 0.60  & 0.16 & 67 & 0.71 & 0.13 & -5.02 & $<$0.001 & -0.75  \\
Self-Efficacy Post & 141 & 0.50 & 0.20 & 68 & 0.63 & 0.19 & -4.60 & $<$0.001 & -0.68  \\
Test Anxiety Pre & 129 & 0.48 & 0.26 & 66 & 0.28 & 0.21 & 5.47 & $<$0.001 & 0.83  \\
Test Anxiety Post & 142 & 0.63 & 0.27 & 68 & 0.40 & 0.27 & 5.58 & $<$0.001 & 0.82  \\ \midrule
Homework & 153 & 88 & 11 & 77 & 87 & 13 & 0.91 & 0.364 & 0.13  \\
Midterm Exams & 153 & 66 & 14 & 77 & 71& 13 &  -2.90& 0.004   & -0.40  \\
Final Exam & 153 & 62 & 13 & 77 & 66 & 14 & -1.95 & 0.052 & -0.27  \\
\bottomrule  
\end{tabular*}}
\end{minipage}
\end{center}
\small\textit{Note.} Sample size, mean, and standard deviation (SD) of prior preparation, motivational factors, and assessment outcomes of students. Results and significance of unpaired $t$-tests are provided. Cohen's \textit{d} effect sizes are also given; a negative \textit{d} indicates that female students had lower scores than male students.
\end{table}

\begin{table}[tb]
\begin{center}
\begin{minipage}{\textwidth}
\tbl{\raggedright Physics 2 prior preparation, self-efficacy, test anxiety, and assessment scores \label{111_table}}
{\begin{tabular*}{\textwidth}{@{\extracolsep{\fill}}lccccccccc@{\extracolsep{\fill}}}
\toprule%
& \multicolumn{3}{@{}c@{}}{Female} & \multicolumn{3}{@{}c@{}}{Male} & \multicolumn{3}{@{}c@{}}{Comparison} \\
\cmidrule{2-4}\cmidrule{5-7}\cmidrule{8-10}%
Variable & N & Mean & SD & N & Mean & SD & \textit{t} & $p$ & $d$\\
\midrule
HS GPA & 117 & 4.18 & 0.41 & 86 & 4.06 & 0.41 & 1.99 & 0.048 & 0.28  \\
SAT Math & 117 & 669 & 71 & 85 & 690 & 66 & -2.04 & 0.042 & -0.29   \\ \midrule
Self-Efficacy Pre & 107 & 0.59 & 0.19 & 77 & 0.65&  0.16& -2.28 & 0.023 &  -0.34 \\
Self-Efficacy Post & 48 & 0.51 & 0.20 & 38 & 0.65 & 0.14 & -3.52 & $<$0.001  & -0.76  \\
Test Anxiety Pre & 105 & 0.56 & 0.26 & 75 & 0.34 & 0.24 & 5.93 & $<$0.001  & 0.90   \\
Test Anxiety Post & 48 & 0.60 & 0.28 & 38 & 0.38 & 0.26 & 3.63 & $<$0.001  & 0.79  \\ \midrule
Homework & 117 & 93 & 15 & 86 & 89& 20 & 1.33 & 0.185  & 0.19  \\
Midterm Exams & 117 & 65 & 12 & 86 &66 & 13 &  -0.64& 0.524 & -0.09  \\
Final Exam & 117 & 55 & 15 & 86 & 55& 17 & -0.24 & 0.811 &  -0.03 \\
\bottomrule
\end{tabular*}}
\end{minipage}
\end{center}
\small\textit{Note.} Sample size, mean, and standard deviation (SD) of prior preparation, motivational factors, and assessment outcomes of students. Results and significance of unpaired $t$-tests are provided. Cohen's \textit{d} effect sizes are also given; a negative \textit{d} indicates that female students had lower scores than male students.
\end{table}

For both Physics 1 and 2, men on average had higher SAT Math scores than women, although~the difference was not statistically significant for Physics 1. Also, women generally had higher high school GPAs than men. All these differences are small to medium ($d\sim 0.2$ to $d\sim 0.5$).  Students were generally well prepared for introductory physics: their average SAT Math scores were well above the 2019 national average of 528~\cite{SAT_scores_2020}, and~average high school GPA was around or above 4.0 on a 5.0 point scale.  High school GPA and SAT Math scores are both shown to be predictors of undergraduate STEM performance~\cite{galla2019}.  However, we cannot assume that men’s higher SAT scores directly translate to physics performance, as~students in this sample show the reverse pattern in calculus 1, with~women having a statistically significantly higher grade than men in this course~\cite{whitcomb2020comparison}.

In general, men reported higher self-efficacy and lower test anxiety than women in both Physics 1 and Physics 2, for~both the pre- and post-surveys. However, the~magnitude of differences differed by construct and course. In~Physics 1, gender differences were large ($d\sim 0.8$) for test anxiety and self-efficacy on both pre- and post-surveys. From~pre to post, the~gender differences remained the same for test anxiety, while they decreased slightly for self-efficacy. 
In Physics 2, gender differences were larger for test anxiety than for self-efficacy on both pre- and post-surveys, with~overall higher self-efficacy and lower test anxiety for men. Moreover, while test anxiety gender gaps decreased from pre to post, there was an increased gender gap for self-efficacy. While men's average self-efficacy score did not change throughout the Physics 2 course, women's self-efficacy score decreased, which contributed to the larger gender~gap.

In terms of broader implications, our results in this novel context of bioscience students in physics courses (in which women outnumber men) are similar to findings for students taking calculus-based introductory physics: there are gender differences in both self-efficacy and test anxiety favoring men~\cite{malespina2022calc_TA}, and~the constructs change over time, showing that they are malleable and potentially able to be influenced~\cite{Malespina2024calc_TA}.

\begin{figure}[h]
    \centering
    \includegraphics[width=0.6\textwidth]{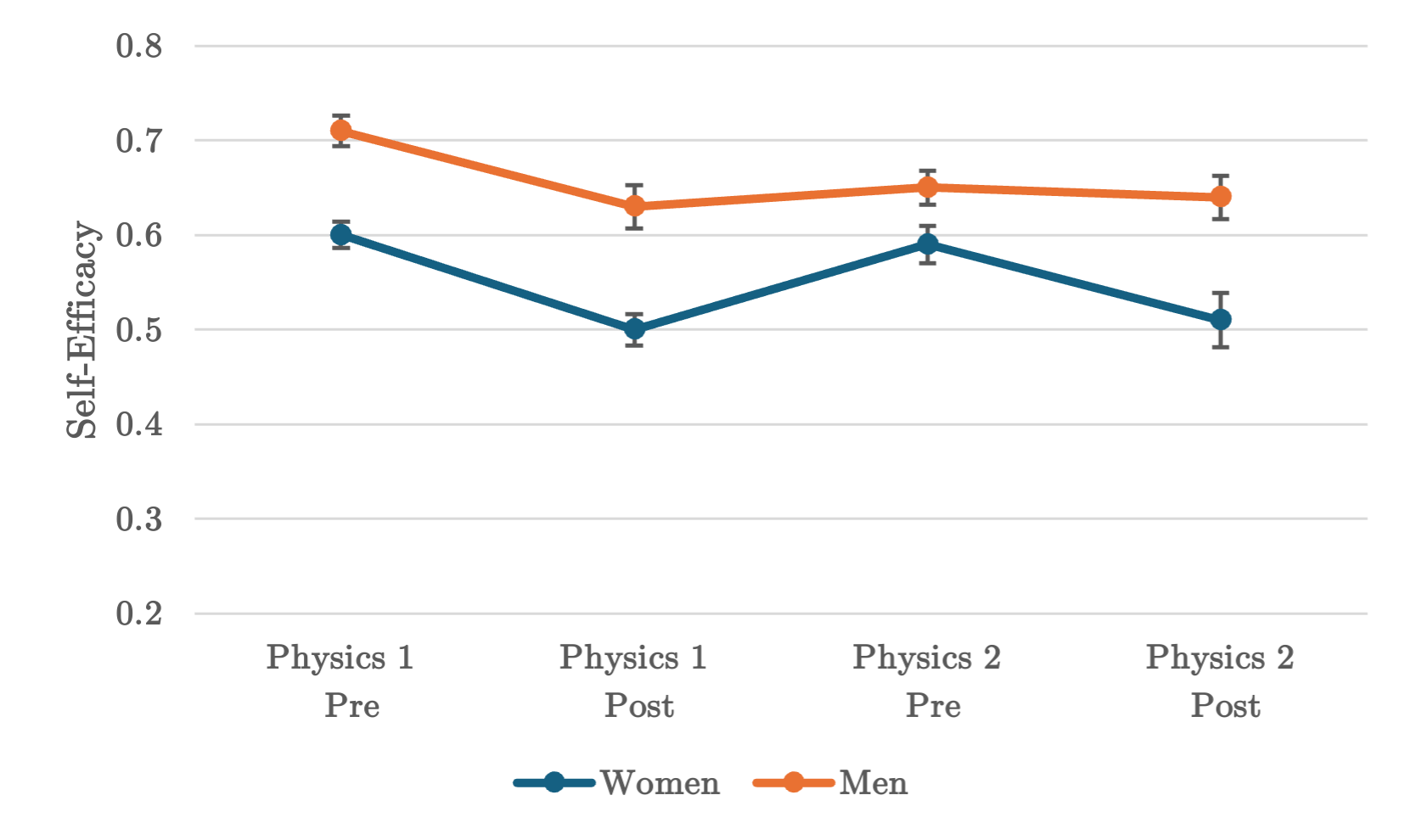}
    \caption{\raggedright Average self-efficacy  scores of men and women from the start of Physics 1 to the end of Physics 2.   Error bars represent standard error and self-efficacy score is on a 0-1 scale. \label{fig:selfeff}}
\end{figure}

\begin{figure}[h]
    \centering
    \includegraphics[width=0.6\textwidth]{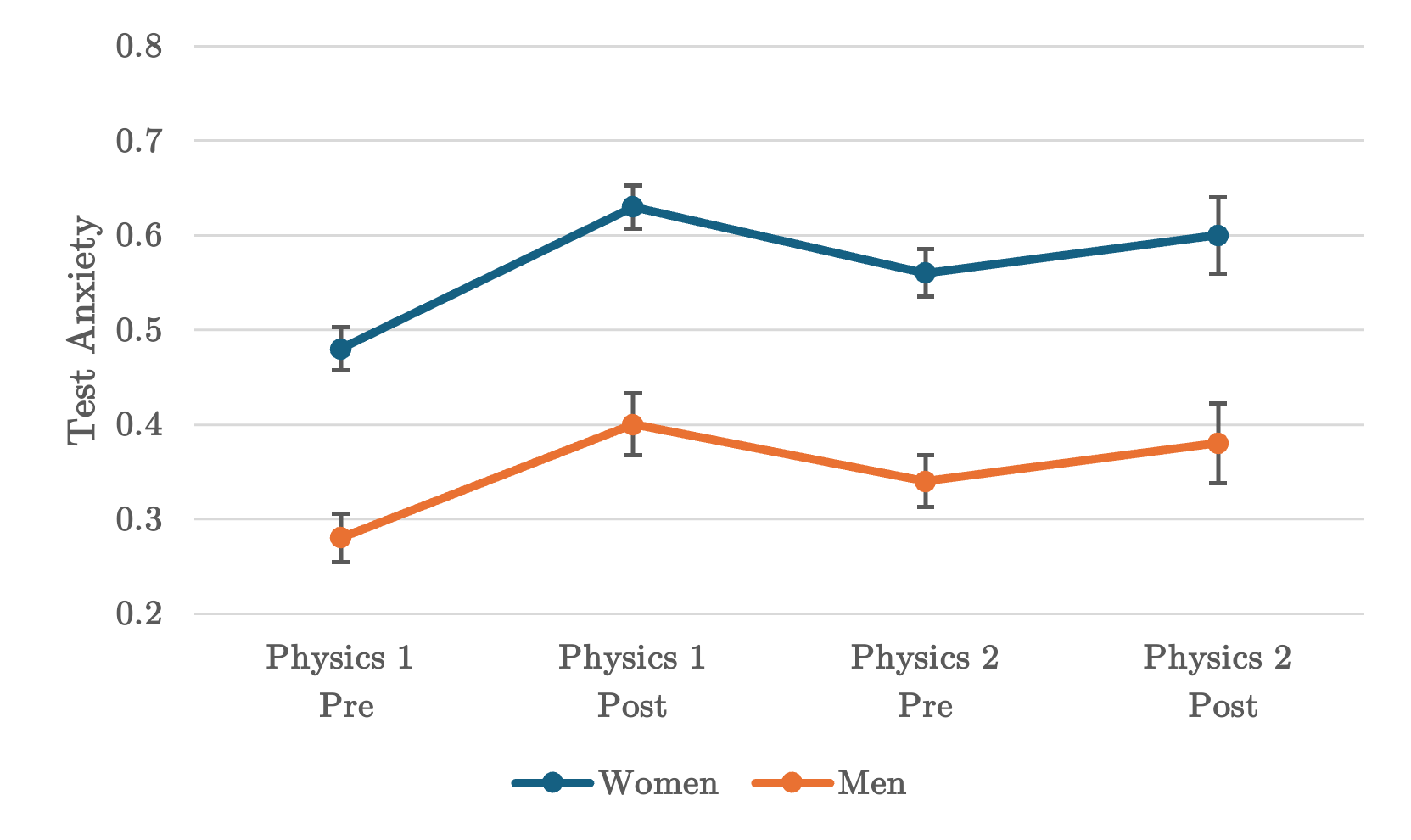}
    \caption{\raggedright Average test anxiety scores of men and women from the start of Physics 1 to the end of Physics 2.  Error bars represent standard error and test anxiety is on a 0-1 scale.  \label{fig:anxiety}}
\end{figure}

We also note differences in motivational factors from Physics 1 to Physics 2. Overall, students' self-efficacy decreased from Physics 1 pre to post, which can be seen in Figure~\ref{fig:selfeff}. For~men, the~average self-efficacy score remained fairly stable from the end of Physics 1 to the end of Physics 2. Overall, there was an increase in self-efficacy and decrease in test anxiety scores for students in our sample going from Physics 1 post to Physics 2 pre. We hypothesize that one of the reasons for this shift may be attributed to students receiving final grades which exceeded their initial expectations based on the average final exams in Table~\ref{110_table}. 
At the start of Physics 2, women reported self-efficacy that was similar to what they reported at the start of Physics 1. However, from~pre to post, their self-efficacy in Physics 2 decreased again, which can also be seen in Figure~\ref{fig:selfeff}.  On~the other hand, Figure~\ref{fig:anxiety} shows that both men and women exhibit a similar trend of increasing test anxiety over the duration of taking Physics 1 and Physics 2 courses, with~women having higher levels of anxiety overall.  A~student affected by test anxiety is likely to experience limits on the cognitive resources they can devote to the assessment~\cite{zeidner1998}, so we hypothesize that increased test anxiety may prevent students from accurately representing their knowledge on high-stakes assessments. In~traditional exam-reliant courses, this is particularly~concerning.

\subsection{RQ2. Are There Gender Differences in Students’ Low- and High-Stakes Assessment Scores?}

Homework constitutes the  ``low-stakes'' assessment category. Tables~\ref{110_table} and \ref{111_table} show that female students had higher average homework scores than male students in Physics 1 (Table \ref{110_table}) and Physics 2 (Table \ref{111_table}). However, both effect sizes were small ($d\sim 0.2$), and gender differences were not statistically significant.  Midterm and final exams constitute ``high-stakes'' assessments.  In~Physics 1, on~average, men had higher exam scores. The~gender differences were statistically significant with a medium effect size for the midterm exams and marginally significant for the final exam. On~the other hand, there were no statistically significant gender differences in exam scores for Physics~2. 

The courses had grades based primarily on midterm and final exam scores. This raises concerns for the Physics 1 course gender differences in exam scores. Research suggests that women are more likely than men to leave STEM fields due to concerns about grades (even if they have a reasonably high average) \cite{seymour2019, rodriguez2016gender}, so women’s lower high-stakes exam scores may contribute to their decision to leave these majors that require introductory physics. This, combined with data that show that gender gaps exist in very few other courses for bioscience students at this institution~\cite{malespina2023biosci_gender}, suggests that  
some of these women may leave bioscience-related STEM fields due to their worse average performance in physics than~men.

Physics 2 shows no such gender disparity in exam scores, though~the course is graded in a similar way to Physics 1. Although~this study is quantitative in nature and does not focus on the mechanisms for this difference, we hypothesize some potential reasons for this difference between Physics 1 and Physics 2. One difference may be that although the percentages of female and male students taking any type of high school physics course are roughly the same and the majority of these courses are heavily focused on Physics 1 content (e.g., mechanics), men are more likely to take AP or honors versions of these high school courses. These AP or honors versions of the high school courses may give men an advantage over women in the college Physics 1 course discussed here and may lead to the gender difference observed in our investigation if the instructor does not provide opportunities to bridge the gender gap due to disparities in high school courses taken~\cite{lindstrom2011self, marchand2013stereotype}. On~the other hand, a~majority of these students do not take high school physics courses with materials commensurate with college Physics 2 (e.g., electricity and magnetism) discussed here. Thus, men are unlikely to have an advantage coming in compared to women. Other potential reasons are that only those who pass Physics 1 with at least a C grade can enroll in Physics 2 and~that while Physics 2 is mandatory for all bioscience majors, it is not mandatory for some health-related majors (although these students from other health-related majors comprise a smaller portion of the students in Physics 1 compared to bioscience majors).

\subsection{RQ3. Are There Gender Differences in Terms of Low- or High-Stakes Assessment Scores for Students with the Same Self-Efficacy and/or Test Anxiety?}

The results of models predicting low-stakes assessments for Physics 1 and 2 can be found in Table~\ref{lowstakes_regressions} in the Appendix. For~both Physics 1 and Physics 2, low-stakes assessment scores are not predicted by pre-self-efficacy, average self-efficacy, pre-test anxiety, or~average test anxiety. Also, the~models that include self-efficacy and test anxiety do not explain any more variance than the models that exclude them. The~only significant predictor of low-stakes assessments is high school GPA, and gender is not a significant predictor of the outcome. This confirms our hypothesis that test anxiety will not predict low-stakes assessment~scores.

Tables~\ref{110_regressions} and \ref{111_regressions} show the results of our  regression models predicting high-stakes assessment outcomes for Physics 1 and Physics 2, respectively. Each of the models uses the variables in the far left column to predict the outcome variable (e.g., Physics 1 and Physics 2 high-stakes assessment scores). Any blank spaces in the table indicate that the predictor in the corresponding row was not used in that model. We have used test anxiety and self-efficacy scores from either the pre-survey or the average score using both pre- and post-survey as predictors. The~adjusted R-squared for each model has been reported as a measure of how well the model explains the variance in the dependent variable while taking the number of predictors into account. The~strength of each predictor, controlling for all other predictors in the model, is given by the standardized regression  coefficient~\cite{cohen2003}. More specifically, for~each change of one standard deviation in the predictor variable, the~model predicts there will be a change of beta standard deviations in the outcome variable, controlling for all other predictor variables~\cite{cohen2003}. The~tables aim to show the predictive relationship between student characteristics (sex, high school GPA, SAT/ACT math scores, self-efficacy, and~test anxiety) and their Physics 1 high-stakes exam scores. The~coefficients, as~described earlier, represent how strongly each characteristic is related to the Physics 1 exam scores, where a positive number indicates a positive relationship and a negative number indicates a negative relationship. For~example, we can see that in  Average Model 1, a~one-standard-deviation increase in test anxiety decreases the Physics 1 high-stakes assessment score by 0.26 standard deviations, when keeping all other variables~constant.

\begin{table}[b]
\begin{center}
\begin{minipage}{\textwidth}
\tbl{\raggedright  Physics 1 high-stakes assessment scores predicted by student sex, High School GPA (HS GPA), SAT/ACT Math scores, pre/average self-efficacy and pre/average test anxiety. \label{110_regressions}}
{\begin{tabular*}{0.8\textwidth}{l|cc|cccc} \toprule
 & \multicolumn{2}{c}{Pre} & \multicolumn{4}{c}{Avg.} \\
Variable & \multicolumn{1}{c}{Model 1} & \multicolumn{1}{c}{Model 2} & \multicolumn{1}{c}{Model 1} & \multicolumn{1}{c}{Model 2} & \multicolumn{1}{c}{Model 3} & \multicolumn{1}{c}{Model 4}  \\ \midrule
Sex & -0.16$^{*}$ & -0.21$^{***}$ & -0.06$^{ns}$ & -0.20$^{***}$  & -0.07$^{ns}$ & -0.10$^{ns}$\\
HS GPA & 0.30$^{***}$ & 0.32$^{***}$ & 0.23$^{***}$ &  0.29$^{***}$&  0.24$^{***}$&  0.25$^{***}$ \\
SAT/ACT Math & 0.41$^{***}$ & 0.41$^{***}$ & 0.37$^{***}$ &  0.42$^{***}$&  0.38$^{***}$ &  0.40$^{***}$\\ \midrule
Self-Efficacy & 0.06$^{ns}$ &  & 0.12$^{ns}$ &   & &0.26$^{***}$ \\   
Test Anxiety & -0.08$^{ns}$ & & -0.26$^{***}$ & & -0.32$^{***}$ & \\\midrule
Adjusted $R^2$ & 0.38 & 0.37 & 0.45 &0.37 & 0.45 & 0.42 \\
N & 190 & 190 & 174 & 174 & 174 & 174\\ \bottomrule
\end{tabular*}}  
\end{minipage}
\end{center}
\bigskip 
\small \textit{Note.} \raggedright Standardized regression ($\beta$) coefficients are provided. $^{*}= p<0.05$, $^{**}= p<0.01$, $^{***}= p<0.001$, and $^{ns}=$ not statistically significant. 

\end{table}

\begin{table}[tb]
\begin{center}
\begin{minipage}{\textwidth}
\tbl{\raggedright  Physics 2 high-stakes assessment scores predicted by student sex, High School GPA (HS GPA), SAT/ACT Math scores, pre/average self-efficacy and pre/average test anxiety. \label{111_regressions}}
{\begin{tabular*}{0.8\textwidth}{l|cc|cccc} \toprule
 & \multicolumn{2}{c}{Pre} & \multicolumn{4}{c}{Avg} \\
Variable & \multicolumn{1}{c}{Model 1} & \multicolumn{1}{c}{Model 2} & \multicolumn{1}{c}{Model 1} & \multicolumn{1}{c}{Model 2} & \multicolumn{1}{c}{Model 3} & \multicolumn{1}{c}{Model 4}  \\ \midrule
Sex & 0.02$^{ns}$ & -0.02$^{ns}$ & 0.09$^{ns}$ & -0.03$^{ns}$ &0.08$^{ns}$  &   0.05$^{ns}$ \\
HS GPA & 0.25$^{***}$ & 0.25$^{***}$ & 0.09$^{ns}$ &  0.09$^{ns}$ &  0.09$^{ns}$&   0.09$^{ns}$\\
SAT/ACT Math & 0.40$^{***}$ & 0.43$^{***}$ & 0.46$^{***}$ & 0.52$^{***}$  & 0.47$^{***}$ &  0.46$^{***}$\\ \midrule
Self-Efficacy & 0.02$^{ns}$ &  & 0.08$^{ns}$ &   & & 0.24$^{*}$\\   
Test Anxiety & -0.11$^{ns}$ & & -0.23$^{ns}$ & & -0.28$^{**}$ & \\\midrule
Adjusted $R^2$ & 0.33 & 0.32 & 0.35 & 0.29 & 0.35 &  0.33\\
N & 179 & 179 & 78 & 78 & 78 & 78\\ \bottomrule
\end{tabular*}}  
\end{minipage}
\end{center}
\bigskip
 \small \textit{Note.} \raggedright Standardized regression ($\beta$) coefficients are provided. $^{*}= p<0.05$, $^{**}= p<0.01$, $^{***}= p<0.001$, and $^{ns}=$ not statistically significant. 
\end{table}

For Physics 1, both high school GPA and SAT/ACT Math scores consistently have strong positive correlations with Physics 1 exam scores across all models. This means that students with higher high school GPAs and higher standardized math scores tend to do better on the high-stakes exams in this course. However, high-stakes assessment scores are not predicted by pre-test anxiety or pre-self-efficacy, which can be seen in Table~\ref{110_regressions}. Also, looking at pre-survey scores, Model 1 in Table~\ref{110_regressions} (which includes pre-self-efficacy and test anxiety as predictors) does not explain much more of the variance than Model 2 (which does not include pre-self-efficacy and test anxiety as predictors). Thus, prior self-efficacy or test-anxiety scores do not appear to greatly contribute to performance  differences among students.  This is good because instructors can intervene by creating an equitable and inclusive learning environment to improve students' self-efficacy and reduce their test~anxiety. 

However, in~Physics 1, high-stakes assessment scores were predicted by average test anxiety but not self-efficacy, which can be seen in Table~\ref{110_regressions}. Average Model 1 in Table~\ref{110_regressions} shows that average test anxiety predicts high-stakes assessment outcomes. Importantly, Average Model 1 and Average Model 3 explain more of the variance in the dependent variable (high-stakes assessment scores) compared to the other models. Additionally, there are no statistically significant gender differences in any of the average models that include test anxiety or self-efficacy as a predictor of high-stakes grades, but~there are gender differences in Average Model 2. This means that gender differences in test anxiety and self-efficacy may account for some of the gender differences disadvantaging women we see in high-stakes~assessments. 

For Physics 2, high-stakes assessment scores were similarly not predicted by pre-self-efficacy or pre-test anxiety, and~there were no gender differences for students with the same prior preparation scores (high school GPA and SAT/ACT math scores). We can see again that the students' prior self-efficacy and test anxiety scores do not seem to contribute to their exam performance, which gives instructors the chance to cultivate a more inclusive and fair learning environment. In~the models using the average score, controlling both for average self-efficacy and test anxiety, neither of them are statistically significant predictors of the outcome, but~test anxiety and self-efficacy each become statistically significant predictors of the high-stakes assessment score in Model 3 and Model 4. This implies that for students in Physics 2, self-efficacy and test anxiety are inversely correlated, meaning that students with high test anxiety tend to have lower self-efficacy and vice versa, which can potentially explain how including both variables obscures their relationship with the outcome. Additionally, gender does not predict high-stakes assessment outcomes in Physics 2. The~models can be seen in Table~\ref{111_regressions}.

Broadly, we find that self-efficacy positively predicts high-stakes assessment scores, while test anxiety negatively predicts scores.  We also find (see Section~\ref{rq1} RQ1) that both self-efficacy and test anxiety measures became worse over time for all students and~that women reported lower self-efficacy and higher test anxiety than men even in this novel context of physics courses in which they outnumber men.  Thus, it is important for instructors to take steps to reduce student test anxiety and increase student self-efficacy. This is important in encouraging the success of all students, but~particularly women who appear to be more affected by low self-efficacy and high test anxiety, especially in Physics~1.  

\section{Conclusions, Limitations, and Future~Directions\label{conclusions}}

We have added to the research in this field by exploring gender differences in and the predictive power of self-efficacy and test anxiety for low- and high-stakes assessments for physics courses in which women are in the majority. While past research has identified gender differences in these factors and their relationship with low- and high-stakes assessment outcomes in courses where women are in the minority~\cite{malespina2022calc_TA}, our results here demonstrate that the inequitable patterns hold even when women outnumber men. Using validated survey data and grade information from students in a two-semester introductory Physics course sequence for bioscience majors and other majors with students interested in health-related professions, we compared the predictive power of self-efficacy and test anxiety on female and male students' performance on both low- and high-stakes assessments. We found that there are gender differences disadvantaging women in self-efficacy, test anxiety, and~high-stakes (but not low-stakes) assessment outcomes in Physics 1. We also found that self-efficacy and/or test anxiety predicted only high-stakes assessment outcomes in both Physics 1 and Physics 2. Comparison of these findings in a novel context with prior studies involving physical science and engineering students shows that although women outnumber men in physics courses for bioscience students and the career goals of bioscience students are very different from those of physical science and engineering majors, most of the adverse trends are similar even for this new population. Therefore, these findings highlight the systemic nature of women being more adversely affected by anxiety in high-stakes assessments and the need for creating more equitable learning~environments.

Instructors can help decrease test anxiety and its adverse impact on student performance directly by lowering the emphasis on high-stakes assessments and increasing the emphasis on low-stakes assessments in their courses.  The~careful choice of assessment tools can help create a more equitable classroom environment by minimizing fears that come with test anxiety related to high-stakes assessments (e.g., receiving a low course grade due to one bad exam score which counts for a significant portion of the course grade). Moreover, frequent low-stakes assessments can also give students many attempts to practice test taking and encourage spaced practice, which is more effective for the retention of knowledge and skill development than ``cramming'' before an exam~\cite{anderson2000}. We recognize that instructors may want to include opportunities to assess students' cumulative learning in a course. In~particular, since  physics is a hierarchical discipline, it may be useful to give students incentives and support to organize their knowledge hierarchically so that they focus on discerning the connections between the concepts in different chapters.  However,   
these types of cumulative assessments can be made lower-stakes by offering more of them in the course and making each count towards less of a student's grade~\cite{laverty2012}. Additionally, implementing a range of formative assessments (such as clicker questions, homework, tutorials, projects, and~other types of assessments), each of which do not count for a very large portion of a students' course grade, can help students develop a wider variety of skills without increasing their anxiety. In~particular, providing students with these types of supports via different types of frequent formative assessments can help students develop the desired knowledge structure and skills including the ability to communicate science better based upon the goals of the course in a low-anxiety and equitable learning~environment.

In addition to recommending the instructors to incorporate more frequent low-stakes assessments to reduce student anxiety and boost self-efficacy, rather than a few heavily weighted high-stakes exams, we also suggest the implementation of research-based interventions and activities designed to promote equity in college courses. An~important implication is that the course instructors should carefully contemplate how to foster equitable, inclusive, and low-anxiety learning environments and examine the role of high-stakes and low-stakes assessments in grading policies.
The course instructors have a central role to play in creating such an environment in which student anxiety is reduced during learning and assessment and self-efficacy is improved~\cite{sathy2022inclusive}. For~example, prior research suggests that when instructors emphasize at the beginning of their courses through short activities that most students struggle in learning to solve challenging physics problems, that struggle is the stepping stone to learning physics, and that students should embrace them, while talking about their own struggles when they were in similar courses, the~gender gap in performance is eliminated~\cite{binning2020belonging}. When instructors implement these types of activities, they are also encouraged to reflect on their own mindset about whether all students can excel in their courses, which is important because instructors with growth mindset about their students' potential have significantly smaller performance differences between traditionally marginalized and dominant demographic groups in STEM courses, and students from traditionally marginalized groups report greater motivation in those courses~\cite{canning2019}.

One limitation of this investigation with regard to RQ3 is that our findings are correlational in nature and a correlation does not imply causation. Another limitation related to generalizability is that the research was carried out at a large research university in the US and results may not apply to other types of US institutions, e.g.,~two-year or four-year liberal arts colleges or higher education institutions in other countries. Moreover, the~institution where this study was carried out is a predominantly White institution, and these findings may not apply to minority-serving institutions. 
Another limitation relates to the fact that the investigation only focused on gender differences and did not focus on other facets of students' identities, e.g.,~based upon their race/ethnicity, first-generation college status, low-income status, etc., and the intersectionality of different identities and how they impact test anxiety and~self-efficacy. 

Future studies could focus on investigating similar issues at other types of institutions as well as for different student demographic groups and intersecting identities. It would also be valuable to conduct individual interviews or focus group discussions with students from various demographic groups because qualitative investigations can complement quantitative study discussed here and shed light on the mechanisms, e.g.,~for the observed gender~differences.


\section*{Authors’ contributions}
A.M. and F.S. contributed to analysis and interpretation of data, as~well as writing and revision of the manuscript. C.S. contributed to the conception and design of research, acquisition and interpretation of data, and~revision of the manuscript.

\section*{Funding}

This research is supported by National Science Foundation grant DUE-1524575.

\section*{Institutional Review}
This research was carried out in accordance with the principles outlined in the university's Institutional Review Board ethical policy with approval code PRO15030412.

\section*{Informed Consent}
Consent requirement was waived due to the research being approved as exempt from informed consent by the university's Institutional Review Board.

\section*{Data Availability Statement}
The datasets used and/or analyzed during the current study are available from the corresponding author on reasonable request.

\section*{Conflicts of Interest}

No potential conflict of interest was reported by the authors.



\bibliographystyle{achemso}
\bibliography{refs}

\providecommand{\latin}[1]{#1}
\makeatletter
\providecommand{\doi}
  {\begingroup\let\do\@makeother\dospecials
  \catcode`\{=1 \catcode`\}=2 \doi@aux}
\providecommand{\doi@aux}[1]{\endgroup\texttt{#1}}
\makeatother
\providecommand*\mcitethebibliography{\thebibliography}
\csname @ifundefined\endcsname{endmcitethebibliography}  {\let\endmcitethebibliography\endthebibliography}{}
\begin{mcitethebibliography}{63}
\providecommand*\natexlab[1]{#1}
\providecommand*\mciteSetBstSublistMode[1]{}
\providecommand*\mciteSetBstMaxWidthForm[2]{}
\providecommand*\mciteBstWouldAddEndPuncttrue
  {\def\EndOfBibitem{\unskip.}}
\providecommand*\mciteBstWouldAddEndPunctfalse
  {\let\EndOfBibitem\relax}
\providecommand*\mciteSetBstMidEndSepPunct[3]{}
\providecommand*\mciteSetBstSublistLabelBeginEnd[3]{}
\providecommand*\EndOfBibitem{}
\mciteSetBstSublistMode{f}
\mciteSetBstMaxWidthForm{subitem}{(\alph{mcitesubitemcount})}
\mciteSetBstSublistLabelBeginEnd
  {\mcitemaxwidthsubitemform\space}
  {\relax}
  {\relax}

\bibitem[Bouffard-Bouchard \latin{et~al.}(1991)Bouffard-Bouchard, Parent, and Larivee]{bouffardbouchard1991}
Bouffard-Bouchard,~T.; Parent,~S.; Larivee,~S. Influence of self-efficacy on self-regulation and performance among junior and senior high-school age students. \emph{International Journal of Behavioral Development} \textbf{1991}, \emph{14}, 153--164\relax
\mciteBstWouldAddEndPuncttrue
\mciteSetBstMidEndSepPunct{\mcitedefaultmidpunct}
{\mcitedefaultendpunct}{\mcitedefaultseppunct}\relax
\EndOfBibitem
\bibitem[Pintrich and De~Groot(1990)Pintrich, and De~Groot]{pintrich1990}
Pintrich,~P.~R.; De~Groot,~E.~V. Motivational and Self-Regulated Learning Components of Classroom Academic Performance. \emph{Journal of Educational Psychology} \textbf{1990}, \emph{82}, 33--40\relax
\mciteBstWouldAddEndPuncttrue
\mciteSetBstMidEndSepPunct{\mcitedefaultmidpunct}
{\mcitedefaultendpunct}{\mcitedefaultseppunct}\relax
\EndOfBibitem
\bibitem[Zimmerman(2000)]{zimmerman2000}
Zimmerman,~B.~J. Self-efficacy: An essential motive to learn. \emph{Contemporary Educational Psychology} \textbf{2000}, \emph{25}, 82--91\relax
\mciteBstWouldAddEndPuncttrue
\mciteSetBstMidEndSepPunct{\mcitedefaultmidpunct}
{\mcitedefaultendpunct}{\mcitedefaultseppunct}\relax
\EndOfBibitem
\bibitem[Nissen and Shemwell(2016)Nissen, and Shemwell]{nissen2016}
Nissen,~J.~M.; Shemwell,~J.~T. Gender, experience, and self-efficacy in introductory physics. \emph{Physical Review Physics Education Research} \textbf{2016}, \emph{12}, 020105\relax
\mciteBstWouldAddEndPuncttrue
\mciteSetBstMidEndSepPunct{\mcitedefaultmidpunct}
{\mcitedefaultendpunct}{\mcitedefaultseppunct}\relax
\EndOfBibitem
\bibitem[Sawtelle \latin{et~al.}(2012)Sawtelle, Brewe, and Kramer]{sawtelle2012}
Sawtelle,~V.; Brewe,~E.; Kramer,~L.~H. Exploring the relationship between self-efficacy and retention in introductory physics. \emph{Journal of Research in Science Teaching} \textbf{2012}, \emph{49}, 1096--1121\relax
\mciteBstWouldAddEndPuncttrue
\mciteSetBstMidEndSepPunct{\mcitedefaultmidpunct}
{\mcitedefaultendpunct}{\mcitedefaultseppunct}\relax
\EndOfBibitem
\bibitem[Steele(1997)]{steele1997threat}
Steele,~C.~M. A threat in the air: How stereotypes shape intellectual identity and performance. \emph{American psychologist} \textbf{1997}, \emph{52}, 613\relax
\mciteBstWouldAddEndPuncttrue
\mciteSetBstMidEndSepPunct{\mcitedefaultmidpunct}
{\mcitedefaultendpunct}{\mcitedefaultseppunct}\relax
\EndOfBibitem
\bibitem[Maries and Singh(2024)Maries, and Singh]{maries2024towards}
Maries,~A.; Singh,~C. Towards meaningful diversity, equity and inclusion in physics learning environments. \emph{Nature Physics} \textbf{2024}, \emph{20}, 367--375\relax
\mciteBstWouldAddEndPuncttrue
\mciteSetBstMidEndSepPunct{\mcitedefaultmidpunct}
{\mcitedefaultendpunct}{\mcitedefaultseppunct}\relax
\EndOfBibitem
\bibitem[Salehi \latin{et~al.}(2019)Salehi, Cotner, Azarin, Carlson, Driessen, Ferry, Harcombe, McGaugh, Wassenberg, Yonas, and Ballen]{salehifrontiers}
Salehi,~S.; Cotner,~S.; Azarin,~S.; Carlson,~E.; Driessen,~M.; Ferry,~V.; Harcombe,~W.; McGaugh,~S.; Wassenberg,~D.; Yonas,~A.; Ballen,~C. Gender Performance Gaps Across Different Assessment Methods and the Underlying Mechanisms: The Case of Incoming Preparation and Test Anxiety. \emph{Frontiers in Education} \textbf{2019}, \emph{4}, Publisher Copyright: {\textcopyright} Copyright {\textcopyright} 2019 Salehi, Cotner, Azarin, Carlson, Driessen, Ferry, Harcombe, McGaugh, Wassenberg, Yonas and Ballen.\relax
\mciteBstWouldAddEndPunctfalse
\mciteSetBstMidEndSepPunct{\mcitedefaultmidpunct}
{}{\mcitedefaultseppunct}\relax
\EndOfBibitem
\bibitem[Marshman \latin{et~al.}(2018)Marshman, Kalender, Nokes-Malach, Schunn, and Singh]{marshman2018femalewithA}
Marshman,~E.~M.; Kalender,~Z.~Y.; Nokes-Malach,~T.; Schunn,~C.; Singh,~C. Female students with A’s have similar physics self-efficacy as male students with C’s in introductory courses: A cause for alarm? \emph{Physical review physics education research} \textbf{2018}, \emph{14}, 020123\relax
\mciteBstWouldAddEndPuncttrue
\mciteSetBstMidEndSepPunct{\mcitedefaultmidpunct}
{\mcitedefaultendpunct}{\mcitedefaultseppunct}\relax
\EndOfBibitem
\bibitem[Cwik and Singh(2022)Cwik, and Singh]{cwik2022gender}
Cwik,~S.; Singh,~C. Gender differences in students’ self-efficacy in introductory physics courses in which women outnumber men predict their grade. \emph{Physical Review Physics Education Research} \textbf{2022}, \emph{18}, 020142\relax
\mciteBstWouldAddEndPuncttrue
\mciteSetBstMidEndSepPunct{\mcitedefaultmidpunct}
{\mcitedefaultendpunct}{\mcitedefaultseppunct}\relax
\EndOfBibitem
\bibitem[Cavallo \latin{et~al.}(2004)Cavallo, Potter, and Rozman]{cavallo2004}
Cavallo,~A.~M.; Potter,~W.~H.; Rozman,~M. Gender differences in learning constructs, shifts in learning constructs, and their relationship to course achievement in a structured inquiry, yearlong college physics course for life science majors. \emph{School Science and Mathematics} \textbf{2004}, \emph{104}, 288--300\relax
\mciteBstWouldAddEndPuncttrue
\mciteSetBstMidEndSepPunct{\mcitedefaultmidpunct}
{\mcitedefaultendpunct}{\mcitedefaultseppunct}\relax
\EndOfBibitem
\bibitem[Zeidner(1998)]{zeidner1998}
Zeidner,~M. \emph{Test Anxiety: {T}he State of the Art}; Springer: New York, New York, 1998\relax
\mciteBstWouldAddEndPuncttrue
\mciteSetBstMidEndSepPunct{\mcitedefaultmidpunct}
{\mcitedefaultendpunct}{\mcitedefaultseppunct}\relax
\EndOfBibitem
\bibitem[Bandura \latin{et~al.}(1999)Bandura, Freeman, and Lightsey]{bandura1999}
Bandura,~A.; Freeman,~W.; Lightsey,~R. Self-efficacy: The exercise of control. \emph{Journal of Cognitive Psychotherapy} \textbf{1999}, \emph{13}, 158--166\relax
\mciteBstWouldAddEndPuncttrue
\mciteSetBstMidEndSepPunct{\mcitedefaultmidpunct}
{\mcitedefaultendpunct}{\mcitedefaultseppunct}\relax
\EndOfBibitem
\bibitem[Maloney \latin{et~al.}(2014)Maloney, Sattizahn, and Beilock]{maloney2014anxiety}
Maloney,~E.~A.; Sattizahn,~J.~R.; Beilock,~S.~L. Anxiety and cognition. \emph{Wiley Interdisciplinary Reviews: Cognitive Science} \textbf{2014}, \emph{5}, 403--411\relax
\mciteBstWouldAddEndPuncttrue
\mciteSetBstMidEndSepPunct{\mcitedefaultmidpunct}
{\mcitedefaultendpunct}{\mcitedefaultseppunct}\relax
\EndOfBibitem
\bibitem[Solomon \latin{et~al.}(1996)Solomon, Battistich, Kim, and Watson]{solomon1996}
Solomon,~D.; Battistich,~V.; Kim,~D.-I.; Watson,~M. Teacher practices associated with students' sense of the classroom as a community. \emph{Social Psychology of Education} \textbf{1996}, \emph{1}, 235--267\relax
\mciteBstWouldAddEndPuncttrue
\mciteSetBstMidEndSepPunct{\mcitedefaultmidpunct}
{\mcitedefaultendpunct}{\mcitedefaultseppunct}\relax
\EndOfBibitem
\bibitem[Ballen \latin{et~al.}(2017)Ballen, Salehi, and Cotner]{ballen2017a}
Ballen,~C.~J.; Salehi,~S.; Cotner,~S. Exams disadvantage women in introductory biology. \emph{PloS One} \textbf{2017}, \emph{12}, e0186419--e0186419\relax
\mciteBstWouldAddEndPuncttrue
\mciteSetBstMidEndSepPunct{\mcitedefaultmidpunct}
{\mcitedefaultendpunct}{\mcitedefaultseppunct}\relax
\EndOfBibitem
\bibitem[Stang \latin{et~al.}(2020)Stang, Altiere, Ives, and Dubois]{stang2020}
Stang,~J.~B.; Altiere,~E.; Ives,~J.; Dubois,~P.~J. Exploring the contributions of self-efficacy and test anxiety to gender differences in assessments. Physics Education Research Conference 2020. Virtual Conference, 2020; pp 497--502\relax
\mciteBstWouldAddEndPuncttrue
\mciteSetBstMidEndSepPunct{\mcitedefaultmidpunct}
{\mcitedefaultendpunct}{\mcitedefaultseppunct}\relax
\EndOfBibitem
\bibitem[Malespina and Singh(2022)Malespina, and Singh]{malespina2022calc_TA}
Malespina,~A.; Singh,~C. Gender differences in test anxiety and self-efficacy: {W}hy instructors should emphasize low-stakes formative assessments in physics courses. \emph{European Journal of Physics} \textbf{2022}, \emph{43}, 035701\relax
\mciteBstWouldAddEndPuncttrue
\mciteSetBstMidEndSepPunct{\mcitedefaultmidpunct}
{\mcitedefaultendpunct}{\mcitedefaultseppunct}\relax
\EndOfBibitem
\bibitem[Bandura(2012)]{bandura_2012}
Bandura,~A. On the Functional Properties of Perceived Self-Efficacy Revisited. \emph{Journal of Management} \textbf{2012}, \emph{38}, 9--44\relax
\mciteBstWouldAddEndPuncttrue
\mciteSetBstMidEndSepPunct{\mcitedefaultmidpunct}
{\mcitedefaultendpunct}{\mcitedefaultseppunct}\relax
\EndOfBibitem
\bibitem[Porter and Ivie(2019)Porter, and Ivie]{porter2019}
Porter,~A.~M.; Ivie,~R. \emph{Women in Physics and Astronomy}; Statistical Research Center of the American Institute of Physics: College Park, MD, 2019\relax
\mciteBstWouldAddEndPuncttrue
\mciteSetBstMidEndSepPunct{\mcitedefaultmidpunct}
{\mcitedefaultendpunct}{\mcitedefaultseppunct}\relax
\EndOfBibitem
\bibitem[Hazari \latin{et~al.}(2017)Hazari, Brewe, Goertzen, and Hodapp]{hazari2017importance}
Hazari,~Z.; Brewe,~E.; Goertzen,~R.~M.; Hodapp,~T. The importance of high school physics teachers for female students’ physics identity and persistence. \emph{The Physics Teacher} \textbf{2017}, \emph{55}, 96--99\relax
\mciteBstWouldAddEndPuncttrue
\mciteSetBstMidEndSepPunct{\mcitedefaultmidpunct}
{\mcitedefaultendpunct}{\mcitedefaultseppunct}\relax
\EndOfBibitem
\bibitem[Li and Singh(2023)Li, and Singh]{li2023impact}
Li,~Y.; Singh,~C. Impact of perceived recognition by physics instructors on women’s self-efficacy and interest. \emph{Physical Review Physics Education Research} \textbf{2023}, \emph{19}, 020125\relax
\mciteBstWouldAddEndPuncttrue
\mciteSetBstMidEndSepPunct{\mcitedefaultmidpunct}
{\mcitedefaultendpunct}{\mcitedefaultseppunct}\relax
\EndOfBibitem
\bibitem[Sundstrom \latin{et~al.}(2022)Sundstrom, Heim, Park, and Holmes]{sundstrom2022introductory}
Sundstrom,~M.; Heim,~A.~B.; Park,~B.; Holmes,~N. Introductory physics students’ recognition of strong peers: Gender and racial or ethnic bias differ by course level and context. \emph{Physical Review Physics Education Research} \textbf{2022}, \emph{18}, 020148\relax
\mciteBstWouldAddEndPuncttrue
\mciteSetBstMidEndSepPunct{\mcitedefaultmidpunct}
{\mcitedefaultendpunct}{\mcitedefaultseppunct}\relax
\EndOfBibitem
\bibitem[Cotner \latin{et~al.}(2020)Cotner, Leno, Walker, J{\o}rgensen, and Vandvik]{cotner2020}
Cotner,~S.; Leno,~L.; Walker,~J.; J{\o}rgensen,~C.; Vandvik,~V. Gender gaps in the performance of {N}orwegian biology students: {T}he roles of test anxiety and science confidence. \emph{International Journal of STEM Education} \textbf{2020}, \emph{7}\relax
\mciteBstWouldAddEndPuncttrue
\mciteSetBstMidEndSepPunct{\mcitedefaultmidpunct}
{\mcitedefaultendpunct}{\mcitedefaultseppunct}\relax
\EndOfBibitem
\bibitem[Pelch(2018)]{pelch2018}
Pelch,~M. Gendered differences in academic emotions and their implications for student success in {STEM}. \emph{International Journal of STEM Education} \textbf{2018}, \emph{5}\relax
\mciteBstWouldAddEndPuncttrue
\mciteSetBstMidEndSepPunct{\mcitedefaultmidpunct}
{\mcitedefaultendpunct}{\mcitedefaultseppunct}\relax
\EndOfBibitem
\bibitem[Crouch and Geller(2023)Crouch, and Geller]{couch2023_incollection}
Crouch,~C.~H.; Geller,~B. In \emph{The International Handbook of Physics Education Research: Learning Physics}; Ta\c{s}ar,~M.~F., Heron,~P. R.~L., Eds.; AIP Publishing: Melville, New York, 2023\relax
\mciteBstWouldAddEndPuncttrue
\mciteSetBstMidEndSepPunct{\mcitedefaultmidpunct}
{\mcitedefaultendpunct}{\mcitedefaultseppunct}\relax
\EndOfBibitem
\bibitem[Malespina(2023)]{Malespinadissertation}
Malespina,~A. Investigating Gender Differences in Test Anxiety, Self-efficacy, Mindset, Grade Penalty and Grades in Physics Courses: A Quest for Equity. Ph.D.\ thesis, University of Pittsburgh, 2023\relax
\mciteBstWouldAddEndPuncttrue
\mciteSetBstMidEndSepPunct{\mcitedefaultmidpunct}
{\mcitedefaultendpunct}{\mcitedefaultseppunct}\relax
\EndOfBibitem
\bibitem[Malespina and Singh(2024)Malespina, and Singh]{Malespina2024calc_TA}
Malespina,~A.; Singh,~C. Introductory physics during COVID-19 remote instruction: Gender gaps in exams are eliminated, but test anxiety and self-efficacy still predict success. \emph{European Journal of Physics} \textbf{2024}, \relax
\mciteBstWouldAddEndPunctfalse
\mciteSetBstMidEndSepPunct{\mcitedefaultmidpunct}
{}{\mcitedefaultseppunct}\relax
\EndOfBibitem
\bibitem[Little \latin{et~al.}(2019)Little, Humphrey, Green, Nair, and Sawtelle]{little2019}
Little,~A.~J.; Humphrey,~B.; Green,~A.; Nair,~A.; Sawtelle,~V. Exploring mindset’s applicability to students’ experiences with challenge in transformed college physics courses. \emph{Physical Review Physics Education Research} \textbf{2019}, \emph{15}, 010127\relax
\mciteBstWouldAddEndPuncttrue
\mciteSetBstMidEndSepPunct{\mcitedefaultmidpunct}
{\mcitedefaultendpunct}{\mcitedefaultseppunct}\relax
\EndOfBibitem
\bibitem[Espinosa \latin{et~al.}(2019)Espinosa, Miller, Araujo, and Mazur]{espinosa2019}
Espinosa,~T.; Miller,~K.; Araujo,~I.; Mazur,~E. Reducing the gender gap in students’ physics self-efficacy in a team-and project-based introductory physics class. \emph{Physical Review Physics Education Research} \textbf{2019}, \emph{15}, 010132\relax
\mciteBstWouldAddEndPuncttrue
\mciteSetBstMidEndSepPunct{\mcitedefaultmidpunct}
{\mcitedefaultendpunct}{\mcitedefaultseppunct}\relax
\EndOfBibitem
\bibitem[Wang and Hazari(2018)Wang, and Hazari]{wang2018}
Wang,~J.; Hazari,~Z. Promoting high school students' physics identity through explicit and implicit recognition. \emph{Physical Review Physics Education Research} \textbf{2018}, \emph{14}, 020111\relax
\mciteBstWouldAddEndPuncttrue
\mciteSetBstMidEndSepPunct{\mcitedefaultmidpunct}
{\mcitedefaultendpunct}{\mcitedefaultseppunct}\relax
\EndOfBibitem
\bibitem[Hazari and Cass(2018)Hazari, and Cass]{hazari2018}
Hazari,~Z.; Cass,~C. Towards meaningful physics recognition: What does this recognition actually look like? \emph{The Physics Teacher} \textbf{2018}, \emph{56}, 442--446\relax
\mciteBstWouldAddEndPuncttrue
\mciteSetBstMidEndSepPunct{\mcitedefaultmidpunct}
{\mcitedefaultendpunct}{\mcitedefaultseppunct}\relax
\EndOfBibitem
\bibitem[Brewe \latin{et~al.}(2010)Brewe, Sawtelle, Kramer, O’Brien, Rodriguez, and Pamelá]{brewe2010}
Brewe,~E.; Sawtelle,~V.; Kramer,~L.~H.; O’Brien,~G.~E.; Rodriguez,~I.; Pamelá,~P. Toward equity through participation in modeling instruction in introductory university physics. \emph{Physical Review Special Topics - Physics Education Research} \textbf{2010}, \emph{6}, 010106\relax
\mciteBstWouldAddEndPuncttrue
\mciteSetBstMidEndSepPunct{\mcitedefaultmidpunct}
{\mcitedefaultendpunct}{\mcitedefaultseppunct}\relax
\EndOfBibitem
\bibitem[Traxler and Brewe(2015)Traxler, and Brewe]{traxler2015}
Traxler,~A.; Brewe,~E. Equity investigation of attitudinal shifts in introductory physics. \emph{Physical Review Special Topics - Physics Education Research} \textbf{2015}, \emph{11}, 020132\relax
\mciteBstWouldAddEndPuncttrue
\mciteSetBstMidEndSepPunct{\mcitedefaultmidpunct}
{\mcitedefaultendpunct}{\mcitedefaultseppunct}\relax
\EndOfBibitem
\bibitem[Van~Dusen and Nissen(2020)Van~Dusen, and Nissen]{vandusen2020_equity}
Van~Dusen,~B.; Nissen,~J. Equity in college physics student learning: A critical quantitative intersectionality investigation. \emph{Journal of Research in Science Teaching} \textbf{2020}, \emph{57}, 33--57\relax
\mciteBstWouldAddEndPuncttrue
\mciteSetBstMidEndSepPunct{\mcitedefaultmidpunct}
{\mcitedefaultendpunct}{\mcitedefaultseppunct}\relax
\EndOfBibitem
\bibitem[Cwik and Singh(2021)Cwik, and Singh]{cwik2021damage}
Cwik,~S.; Singh,~C. Damage caused by societal stereotypes: Women have lower physics self-efficacy controlling for grade even in courses in which they outnumber men. \emph{Physical Review Physics Education Research} \textbf{2021}, \emph{17}, 020138\relax
\mciteBstWouldAddEndPuncttrue
\mciteSetBstMidEndSepPunct{\mcitedefaultmidpunct}
{\mcitedefaultendpunct}{\mcitedefaultseppunct}\relax
\EndOfBibitem
\bibitem[Leslie \latin{et~al.}(2015)Leslie, Cimpian, Meyer, and Freeland]{leslie2015}
Leslie,~S.-J.; Cimpian,~A.; Meyer,~M.; Freeland,~E. Expectations of brilliance underlie gender distributions across academic disciplines. \emph{Science} \textbf{2015}, \emph{347}, 262--265\relax
\mciteBstWouldAddEndPuncttrue
\mciteSetBstMidEndSepPunct{\mcitedefaultmidpunct}
{\mcitedefaultendpunct}{\mcitedefaultseppunct}\relax
\EndOfBibitem
\bibitem[Whitcomb \latin{et~al.}(2020)Whitcomb, Kalender, Nokes-Malach, Schunn, and Singh]{whitcomb2020comparison}
Whitcomb,~K.~M.; Kalender,~Z.~Y.; Nokes-Malach,~T.~J.; Schunn,~C.~D.; Singh,~C. Comparison of self-efficacy and performance of engineering undergraduate women and men. \emph{International Journal of Engineering Education} \textbf{2020}, \emph{36}, 1996--2014\relax
\mciteBstWouldAddEndPuncttrue
\mciteSetBstMidEndSepPunct{\mcitedefaultmidpunct}
{\mcitedefaultendpunct}{\mcitedefaultseppunct}\relax
\EndOfBibitem
\bibitem[Witherspoon and Schunn(2020)Witherspoon, and Schunn]{witherspoon2020locating}
Witherspoon,~E.~B.; Schunn,~C.~D. Locating and understanding the largest gender differences in pathways to science degrees. \emph{Science Education} \textbf{2020}, \emph{104}, 144--163\relax
\mciteBstWouldAddEndPuncttrue
\mciteSetBstMidEndSepPunct{\mcitedefaultmidpunct}
{\mcitedefaultendpunct}{\mcitedefaultseppunct}\relax
\EndOfBibitem
\bibitem[Traxler \latin{et~al.}(2016)Traxler, Cid, Blue, and Barthelemy]{traxler2016}
Traxler,~A.~L.; Cid,~X.~C.; Blue,~J.; Barthelemy,~R. Enriching gender in physics education research: A binary past and a complex future. \emph{Physical Review Physics Education Research} \textbf{2016}, \emph{12}, 020114\relax
\mciteBstWouldAddEndPuncttrue
\mciteSetBstMidEndSepPunct{\mcitedefaultmidpunct}
{\mcitedefaultendpunct}{\mcitedefaultseppunct}\relax
\EndOfBibitem
\bibitem[Pintrich(1991)]{pintrich1991}
Pintrich,~P.~R. \emph{A Manual for the Use of the Motivated Strategies for Learning Questionnaire}; University of Michigan: Ann Arbor, Michigan, 1991\relax
\mciteBstWouldAddEndPuncttrue
\mciteSetBstMidEndSepPunct{\mcitedefaultmidpunct}
{\mcitedefaultendpunct}{\mcitedefaultseppunct}\relax
\EndOfBibitem
\bibitem[Pintrich \latin{et~al.}(1993)Pintrich, Smith, Garcia, and McKeachie]{pintrich_valid}
Pintrich,~P.~R.; Smith,~D. A.~F.; Garcia,~T.; McKeachie,~W.~J. Reliability and Predictive Validity of the Motivated Strategies for Learning Questionnaire ({MSLQ}). \emph{Educational and Psychological Measurement} \textbf{1993}, \emph{53}, 801--813\relax
\mciteBstWouldAddEndPuncttrue
\mciteSetBstMidEndSepPunct{\mcitedefaultmidpunct}
{\mcitedefaultendpunct}{\mcitedefaultseppunct}\relax
\EndOfBibitem
\bibitem[Kalender \latin{et~al.}(2019)Kalender, Marshman, Schunn, Nokes-Malach, and Singh]{kalender2019gendered}
Kalender,~Z.~Y.; Marshman,~E.; Schunn,~C.~D.; Nokes-Malach,~T.~J.; Singh,~C. Gendered patterns in the construction of physics identity from motivational factors. \emph{Physical Review Physics Education Research} \textbf{2019}, \emph{15}, 020119\relax
\mciteBstWouldAddEndPuncttrue
\mciteSetBstMidEndSepPunct{\mcitedefaultmidpunct}
{\mcitedefaultendpunct}{\mcitedefaultseppunct}\relax
\EndOfBibitem
\bibitem[Hu and Bentler(1999)Hu, and Bentler]{hu1999}
Hu,~L.; Bentler,~P.~M. Cutoff criteria for fit indexes in covariance structure analysis: Conventional criteria versus new alternatives. \emph{Structural Equation Modeling: A Multidisciplinary Journal} \textbf{1999}, \emph{6}, 1--55\relax
\mciteBstWouldAddEndPuncttrue
\mciteSetBstMidEndSepPunct{\mcitedefaultmidpunct}
{\mcitedefaultendpunct}{\mcitedefaultseppunct}\relax
\EndOfBibitem
\bibitem[Browne and Cudeck(1993)Browne, and Cudeck]{browne1993}
Browne,~M.~W.; Cudeck,~R. Alternative ways of assessing model fit. \emph{Sage Focus Editions} \textbf{1993}, \emph{154}, 136--136\relax
\mciteBstWouldAddEndPuncttrue
\mciteSetBstMidEndSepPunct{\mcitedefaultmidpunct}
{\mcitedefaultendpunct}{\mcitedefaultseppunct}\relax
\EndOfBibitem
\bibitem[Frey(2018)]{frey2018}
Frey,~B.~B. \emph{The {SAGE} {E}ncyclopedia of {E}ducational {R}esearch, {M}easurement, and {E}valuation}; SAGE Publications, Inc.: Thousand Oaks, CA, 2018\relax
\mciteBstWouldAddEndPuncttrue
\mciteSetBstMidEndSepPunct{\mcitedefaultmidpunct}
{\mcitedefaultendpunct}{\mcitedefaultseppunct}\relax
\EndOfBibitem
\bibitem[Galla \latin{et~al.}(2019)Galla, Shulman, Plummer, Gardner, Hutt, Goyer, D’Mello, Finn, and Duckworth]{galla2019}
Galla,~B.~M.; Shulman,~E.~P.; Plummer,~B.~D.; Gardner,~M.; Hutt,~S.~J.; Goyer,~J.~P.; D’Mello,~S.~K.; Finn,~A.~S.; Duckworth,~A.~L. Why High School Grades Are Better Predictors of On-Time College Graduation Than Are Admissions Test Scores: The Roles of Self-Regulation and Cognitive Ability. \emph{American Educational Research Journal} \textbf{2019}, \emph{56}, 2077--2115\relax
\mciteBstWouldAddEndPuncttrue
\mciteSetBstMidEndSepPunct{\mcitedefaultmidpunct}
{\mcitedefaultendpunct}{\mcitedefaultseppunct}\relax
\EndOfBibitem
\bibitem[Hazari \latin{et~al.}(2007)Hazari, Tai, and Sadler]{hazari2007}
Hazari,~Z.; Tai,~R.~H.; Sadler,~P.~M. Gender differences in introductory university physics performance: {T}he influence of high school physics preparation and affective factors. \emph{Science Education} \textbf{2007}, \emph{91}, 847--876\relax
\mciteBstWouldAddEndPuncttrue
\mciteSetBstMidEndSepPunct{\mcitedefaultmidpunct}
{\mcitedefaultendpunct}{\mcitedefaultseppunct}\relax
\EndOfBibitem
\bibitem[Vincent-Ruz \latin{et~al.}(2018)Vincent-Ruz, Binning, Schunn, and Grabowski]{vincentruz2017}
Vincent-Ruz,~P.; Binning,~K.; Schunn,~C.; Grabowski,~J. The effect of math {SAT} on women's chemistry competency beliefs. \emph{Chemistry Education Research and Practice} \textbf{2018}, \emph{19}, 342--351\relax
\mciteBstWouldAddEndPuncttrue
\mciteSetBstMidEndSepPunct{\mcitedefaultmidpunct}
{\mcitedefaultendpunct}{\mcitedefaultseppunct}\relax
\EndOfBibitem
\bibitem[{The College Board} and {ACT}(2018){The College Board}, and {ACT}]{collegeboard2018}
{The College Board}; {ACT} {G}uide to the 2018 {ACT}/{SAT} {C}oncordance. 2018; \url{https://www.act.org/content/dam/act/unsecured/documents/ACT-SAT-Concordance-Information.pdf}\relax
\mciteBstWouldAddEndPuncttrue
\mciteSetBstMidEndSepPunct{\mcitedefaultmidpunct}
{\mcitedefaultendpunct}{\mcitedefaultseppunct}\relax
\EndOfBibitem
\bibitem[Cohen(1988)]{cohen1988}
Cohen,~J. \emph{Statistical Power Analysis for the Behavioral Sciences}; L. Erlbaum Associates: Hillsdale, N.J., 1988\relax
\mciteBstWouldAddEndPuncttrue
\mciteSetBstMidEndSepPunct{\mcitedefaultmidpunct}
{\mcitedefaultendpunct}{\mcitedefaultseppunct}\relax
\EndOfBibitem
\bibitem[Cohen \latin{et~al.}(2003)Cohen, Cohen, West, and Aiken]{cohen2003}
Cohen,~J.; Cohen,~P.; West,~S.; Aiken,~L.~S. \emph{Applied Multiple Regression/Correlation Analysis for the Behavioral Sciences}; Lawrence Erlbaum, 2003\relax
\mciteBstWouldAddEndPuncttrue
\mciteSetBstMidEndSepPunct{\mcitedefaultmidpunct}
{\mcitedefaultendpunct}{\mcitedefaultseppunct}\relax
\EndOfBibitem
\bibitem[{College Entrance Examination Board}(2020)]{SAT_scores_2020}
{College Entrance Examination Board} {SAT} {S}uite of {A}ssessments {A}nnual {R}eport. 2020; \url{https://research.collegeboard.org/programs/sat/data/2019-sat-suite-annual-report}\relax
\mciteBstWouldAddEndPuncttrue
\mciteSetBstMidEndSepPunct{\mcitedefaultmidpunct}
{\mcitedefaultendpunct}{\mcitedefaultseppunct}\relax
\EndOfBibitem
\bibitem[Seymour \latin{et~al.}(2019)Seymour, Hunter, Thiry, Weston, Harper, Holland, Koch, and Drake]{seymour2019}
Seymour,~E.; Hunter,~A.-B.; Thiry,~H.; Weston,~T.~J.; Harper,~R.~P.; Holland,~D.~G.; Koch,~A.~K.; Drake,~B.~M. \emph{Talking {A}bout {L}eaving {R}evisited: Persistence, {R}elocation, and {L}oss in {U}ndergraduate {STEM} {E}ducation}; Springer International Publishing AG: Cham, 2019\relax
\mciteBstWouldAddEndPuncttrue
\mciteSetBstMidEndSepPunct{\mcitedefaultmidpunct}
{\mcitedefaultendpunct}{\mcitedefaultseppunct}\relax
\EndOfBibitem
\bibitem[Rodriguez \latin{et~al.}(2016)Rodriguez, Potvin, and Kramer]{rodriguez2016gender}
Rodriguez,~I.; Potvin,~G.; Kramer,~L.~H. How gender and reformed introductory physics impacts student success in advanced physics courses and continuation in the physics major. \emph{Physical Review Physics Education Research} \textbf{2016}, \emph{12}, 020118\relax
\mciteBstWouldAddEndPuncttrue
\mciteSetBstMidEndSepPunct{\mcitedefaultmidpunct}
{\mcitedefaultendpunct}{\mcitedefaultseppunct}\relax
\EndOfBibitem
\bibitem[Malespina and Singh(2023)Malespina, and Singh]{malespina2023biosci_gender}
Malespina,~A.; Singh,~C. Gender gaps in grades versus grade penalties: Why grade anomalies may be more detrimental for women aspiring for careers in biological sciences. \emph{International Journal of STEM Education} \textbf{2023}, \emph{10}, 13\relax
\mciteBstWouldAddEndPuncttrue
\mciteSetBstMidEndSepPunct{\mcitedefaultmidpunct}
{\mcitedefaultendpunct}{\mcitedefaultseppunct}\relax
\EndOfBibitem
\bibitem[Lindstr{\o}m and Sharma(2011)Lindstr{\o}m, and Sharma]{lindstrom2011self}
Lindstr{\o}m,~C.; Sharma,~M.~D. Self-efficacy of first year university physics students: Do gender and prior formal instruction in physics matter? \emph{International Journal of Innovation in Science and Mathematics Education} \textbf{2011}, \emph{19}\relax
\mciteBstWouldAddEndPuncttrue
\mciteSetBstMidEndSepPunct{\mcitedefaultmidpunct}
{\mcitedefaultendpunct}{\mcitedefaultseppunct}\relax
\EndOfBibitem
\bibitem[Marchand and Taasoobshirazi(2013)Marchand, and Taasoobshirazi]{marchand2013stereotype}
Marchand,~G.~C.; Taasoobshirazi,~G. Stereotype threat and women's performance in physics. \emph{International Journal of Science Education} \textbf{2013}, \emph{35}, 3050--3061\relax
\mciteBstWouldAddEndPuncttrue
\mciteSetBstMidEndSepPunct{\mcitedefaultmidpunct}
{\mcitedefaultendpunct}{\mcitedefaultseppunct}\relax
\EndOfBibitem
\bibitem[Anderson(2000)]{anderson2000}
Anderson,~J.~R. \emph{Learning and {M}emory: An {I}ntegrated {A}pproach}; Wiley: New York, 2000\relax
\mciteBstWouldAddEndPuncttrue
\mciteSetBstMidEndSepPunct{\mcitedefaultmidpunct}
{\mcitedefaultendpunct}{\mcitedefaultseppunct}\relax
\EndOfBibitem
\bibitem[Laverty \latin{et~al.}(2012)Laverty, Bauer, Kortemeyer, and Westfall]{laverty2012}
Laverty,~J.~T.; Bauer,~W.; Kortemeyer,~G.; Westfall,~G. Want to Reduce Guessing and Cheating While Making Students Happier? {G}ive More Exams! \emph{The Physics Teacher} \textbf{2012}, \emph{50}, 540--543\relax
\mciteBstWouldAddEndPuncttrue
\mciteSetBstMidEndSepPunct{\mcitedefaultmidpunct}
{\mcitedefaultendpunct}{\mcitedefaultseppunct}\relax
\EndOfBibitem
\bibitem[Sathy and Hogan(2022)Sathy, and Hogan]{sathy2022inclusive}
Sathy,~V.; Hogan,~K.~A. \emph{Inclusive teaching: Strategies for promoting equity in the college classroom}; West Virginia University Press, 2022\relax
\mciteBstWouldAddEndPuncttrue
\mciteSetBstMidEndSepPunct{\mcitedefaultmidpunct}
{\mcitedefaultendpunct}{\mcitedefaultseppunct}\relax
\EndOfBibitem
\bibitem[Binning \latin{et~al.}(2020)Binning, Kaufmann, McGreevy, Fotuhi, Chen, Marshman, Kalender, Limeri, Betancur, and Singh]{binning2020belonging}
Binning,~K.~R.; Kaufmann,~N.; McGreevy,~E.~M.; Fotuhi,~O.; Chen,~S.; Marshman,~E.; Kalender,~Z.~Y.; Limeri,~L.; Betancur,~L.; Singh,~C. Changing social contexts to foster equity in college science courses: An ecological-belonging intervention. \emph{Psychological Science} \textbf{2020}, \emph{31}, 1059--1070\relax
\mciteBstWouldAddEndPuncttrue
\mciteSetBstMidEndSepPunct{\mcitedefaultmidpunct}
{\mcitedefaultendpunct}{\mcitedefaultseppunct}\relax
\EndOfBibitem
\bibitem[Canning \latin{et~al.}(2019)Canning, Muenks, Green, and Murphy]{canning2019}
Canning,~E.~A.; Muenks,~K.; Green,~D.~J.; Murphy,~M.~C. {STEM} faculty who believe ability is fixed have larger racial achievement gaps and inspire less student motivation in their classes. \emph{Science Advances} \textbf{2019}, \emph{5}, eaau4734\relax
\mciteBstWouldAddEndPuncttrue
\mciteSetBstMidEndSepPunct{\mcitedefaultmidpunct}
{\mcitedefaultendpunct}{\mcitedefaultseppunct}\relax
\EndOfBibitem
\end{mcitethebibliography}

\newpage

\appendix
\section*{Appendix}

\begin{table*}[ht]
\begin{center}
\begin{minipage}{\textwidth}
\tbl{\raggedright Survey fit indices for confirmatory factor analysis for both Physics 1 and Physics 2.  \label{fit_indicies}}
{\begin{tabular*}{0.9\textwidth}{l|cccc|cccc}
\toprule%
& \multicolumn{4}{@{}c@{}}{Fit Indicies} & \multicolumn{4}{@{}c@{}}{Cronbach's $\alpha$} \\
Course & CFI & TLI & RMSEA & SRMR & Pre SE & Post SE & Pre TA & Post TA\\ \midrule
Physics 1 & 0.95 &  0.94  & 0.07  & 0.05 & 0.71 &  0.81 & 0.89 & 0.92 \\
Physics 2  & 0.93 &  0.91  & 0.08 & 0.06 & 0.84 & 0.87 & 0.93 & 0.94 \\
\bottomrule
\end{tabular*}}
\end{minipage}
\end{center}
\small \textit{Note.} Chronbach's $\alpha$ for pre and post Test Anxiety (TA) and Self-Efficacy (SE) is included. For Physics 1, $N=230$ and for Physics 2, $N=203$. 
\end{table*}

\begin{table*}[htb]
\begin{center}
\begin{minipage}{\textwidth}
\tbl{\raggedright  Physics 1 and 2 low-stakes assessment scores predicted by student sex, High School GPA (HS GPA), SAT/ACT Math scores, pre or average self-efficacy and pre or average test anxiety. \label{lowstakes_regressions}}
{\begin{tabular*}{\textwidth}{l|ccc|c|ccc|c} \toprule
 & \multicolumn{4}{c|}{Physics 1} & \multicolumn{4}{c}{Physics 2} \\
 & \multicolumn{3}{c}{Pre} & \multicolumn{1}{c|}{Avg.} & \multicolumn{3}{c}{Pre} & \multicolumn{1}{c}{Avg.}  \\ \midrule
Variable & Model 1 & Model 2 & Model 3 & Model 1 & Model 1 & Model 2 & Model 3 & Model 1 \\ \midrule
Sex (M=0, F=1) & -0.01$^{ns}$ & -0.02$^{ns}$ & -0.01$^{ns}$& -0.04$^{ns}$ & 0.04$^{ns}$ & 0.04$^{ns}$ & 0.08$^{ns}$ & -0.04$^{ns}$  \\
HS GPA & 0.32$^{***}$ & 0.32$^{***}$ & 0.31$^{***}$&  0.31$^{***}$ & 0.27$^{***}$ & 0.27$^{***}$ & 0.27$^{***}$ & 0.32$^{**}$ \\ 
SAT/ACT Math & 0.05$^{ns}$ & 0.05$^{ns}$ & 0.05$^{ns}$ & 0.05$^{ns}$ & -0.06$^{ns}$ & -0.06$^{ns}$ & -0.09$^{ns}$ & -0.02$^{ns}$   \\\midrule
Self-Efficacy & 0.05$^{ns}$ & & & -0.05$^{ns}$ & -0.02$^{ns}$ &  &  & 0.09$^{ns}$  \\   
Test Anxiety & 0.07$^{ns}$ & 0.04$^{ns}$& & 0.07$^{ns}$ & 0.09$^{ns}$ & 0.10$^{ns}$ &  & 0.04$^{ns}$   \\\midrule
N & 190 & 190& 190& 174 & 179 & 179 & 179 & 78 \\
Adjusted $R^2$ & 0.08& 0.09& 0.09 & 0.07 & 0.06 & 0.06 & 0.06 & 0.04 \\ \bottomrule
\end{tabular*}}
\end{minipage}
\end{center}
\small \textit{Note.} Standardized regression ($\beta$) coefficients are provided. $^{*}= p<0.05$, $^{**}= p<0.01$, $^{***}= p<0.001$, and $^{ns}=$ not statistically significant.  
\end{table*}

\end{document}